\documentclass{czjphys}         
\usepackage{graphicx}
\begin{document}

\title{LO-phonon overheating in quantum dots}  
%
\authori{Karel Kr\'al\footnote{kral@fzu.cz}}      \addressi{Institute of Physics,
Academy of Sciences of Czech Republic, \\ Na Slovance 2, 18221 Prague,
Czech Republic}
\authorii{}     \addressii{}
\authoriii{}    \addressiii{}
\authoriv{}     \addressiv{}
\authorv{}      \addressv{}
\authorvi{}     \addressvi{}
%
\headauthor{K. Kr\'al }
\headtitle{LO-phonon overheating \ldots}  
\lastevenhead{K. Kr\'al: LO-phonon overheating \ldots}
\refnum{A}
\daterec{XXX}    
\issuenumber{0}  \year{2006}
\setcounter{page}{1}
\maketitle

\begin{abstract}
Longitudinal optical phonons have been used to interpret the electronic
energy relaxation in quantum dots and at the same time they served as a
reservoir, with which the electronic subsystem is in contact. Such a
phonon subsystem is expected to be passive, namely,  in a long-time
limit the whole system should be able to achieve such a stationary
state, in which statistical distributions of both subsystems do not
change in time. We pay attention to this property of the LO phonon
bath. We show the passivity property of the so far used approximations
to electronic transport in quantum dots. Also we show a way how to
improve the passivity of LO phonon bath using canonical Lang-Firsov
transformation.
\end{abstract}

\section{Introduction}
Semiconductor nanostructures, like quantum dots, appear to bring new
properties  due to their restricted size in all three dimensions. The
size of them is often not much larger that the charge carrier's mean
free path. Electronic energy relaxation in quantum dots does not seem a
fully resolved  question, both in experiment and theory \cite{yoffe}.
The rapidity of electronic relaxation may be ascribed to the coupling
of electrons with longitudinal optical (LO) phonons
\cite{japss1998,tsuchiya}, although the relaxation
effect in quantum dots was also interpreted with help of Auger
mechanism. Besides this, theory of polarons in quantum dots has
developed recently, which concept, together with the finite lifetime of
LO phonons, has became another way of explanation of the rapid
relaxation of electronic energy in quantum dots
\cite{verzelen2}.

The quantum transport theory, build\-ing in the concept of
non\-equilibrium Gre\-en's func\-tions \cite{LL10}, or alternatively on
using the theory of nonequilibrium statistical operator \cite{zubarev},
used the self-consistent Born approximation to the electronic
self-energy. This approximation to the quantum transport of charge
carriers in quantum dots was able to provide explanation of the
rapidity of electronic relaxation
\cite{japss1998,tsuchiya}, and of some properties of the optical
transitions, like the line-shape \cite{japrb5}, but it also recently
provided arguments in favor of electronic level occupation
up-conversion and incomplete depopulation in quantum dots, which effect
appears in remarkable agreement with a number of experimental papers
(see e.g. \cite{jaIEEE2004,jaSS2004}). The self-consistent Born (SCB)
approximation to electronic self-energy, which we are going to use
here, means that the electron is bound to a cloud of virtual LO
phonons. It is already at this level that we consider explicitly the
motion of the reservoir and its entanglement with the electronic
subsystem. This optical phonon cloud can be viewed as coherent phonons
coupled to an electron. In analogy with coherent light, this coherent
phonons may act as an effective classical time-dependent field, in
which the electron moves. Due to this feature we can meet with the
off-shell property of the collision integral of the electronic quantum
transport equation in the SCB approximation.

Because this self-consistent Born approximation to the electron quantum
transport equation leads to an off-shell formulation of the transport
equation, a question naturally arises concerning an overall energetic
balance in the processes  of electronic energy relaxation via
electron-LO-phonon scattering in quantum dots. Within this subject we
can raise  questions about the passivity of the phonon reservoir. In
particular, we may ask, whether the whole system of electrons and
phonons, can evolve to a long-time limit, such that in this limit the
density matrices of both subsystems are in a steady state. If this
happens to be the case, we can speak about a passivity of the phonon
reservoir. Naturally, this passivity property is a desired
characteristics of the theory describing the system under
consideration, while the opposite case means a difficulty.

The questions of passivity is met in the Gaussian white noise problem
\cite{haug-jauhoQKTES} and in the theory of interacting harmonic oscillators
\cite{Nieuwenhuizen-Allahverdyan,Caldeira-Leggett}.
In this work we demonstrate, using quantum transport equations, that
the problem of a non-passive reservoir occurs in quantum dots. We shall
identify a significant source of such non-passivity with certain terms
in the electron-phonon interaction operator of the system. These terms
are those, which play a leading role in the creation of the polaron
well of the charge carrier in a dot. The existence of the overheating
in quantum dots shows that the formulation of the phonon kinetic
equation need to be approached with care. We shall show a possibility
of a partial elimination of this effect. This will be done by changing
the basis set of electron and phonon eigenstates, switching from the
basis from noninteracting electron and phonons to a set of dressed
states of electron and phonons, which assumes that the crystal lattice
accommodates to the presence of electrons in quantum dot. The
transition to the new dressed basis is achieved by performing the
well-known canonical Lang-Firsov transformation \cite{mahan} of the
Hamiltonian of the system, which exactly diagonalizes a certain part of
the full electron-phonon Hamiltonian, including those terms of the
electron-phonon coupling, which lead to a formation of the electronic
polaron well. The result of the transformation then is a Hamiltonian
for new quasiparticles interacting by a new interaction operator. The
advantage of this new picture then is that the new quasiparticles do
not tend to develop quickly a polaron potential well around them. The
phonon kinetic equation may then become free of corresponding spurious
effects like an excess heating of phonons.

Considering electronic relaxation and phonon generation in the system
of a single quantum dot with a single spin-less electron and with the
influence of holes in the valence band states of quantum dot completely
neglected
\cite{japss1998,jaIEEE2004,jaSS2004},  we shall demonstrate the existence of the
phonon overheating effect in quantum dots and show how to eliminate it
to a considerable degree. We also show that the electronic relaxation
characteristics, given by the electronic transport equation, remain
nearly unchanged by the treatment of the overheating effect with help
of the Lang-Firsov transformation.

\section{Hamiltonian}         
 The electronic eigenstates of this single electron in a quantum
dot will be approximated by electronic eigenstates in a cubic shaped
quantum dot, with infinitely deep potential. The present model assumes
only two nondegenerate electron states $\psi_n({\bf r})$ in this dot,
namely the ground state ($n$=0) and one of the lowest-energy
triple-degenerate exited states ($n$=1). The bath of LO phonons will be
approximated by the bulk modes of LO phonons of the full sample. The
Hamiltonian of the whole system is:
\be
H =H_e+H_{ph}+H_1,
\label{original}
\ee
in which $ H_e=\sum_{i=0,1}E_ic^+_ic_i$, with $c_i$ being annihilation
operator of electron in state $i$. We set the energy of the ground
state $E_0=0$. $ H_{ph}=\sum_{\bf q}E_{LO}b^+_{\bf q}b_{\bf q}$, with
$b_{\bf q}$ being LO phonon annihilation operator in state with phonon
momentum ${\bf q}$. The interaction of electron with modes ${\bf q}$ of
phonons, with phonon energy $E_{LO}$, is
\be
H_1=\sum_{{\bf q},  m,n=0,1} A_q\Phi(n,m,{\bf q})(b_{\bf q}- b^+_{-{\bf
q}})c^+_nc_m.
\ee
The Fr\"ohlich's coupling $H_1$ \cite{callaway} between the electron
and LO phonons contains the coupling constant $A_q$, which is $A_q=(-i
e/q)[E_{LO} (\kappa^{-1}_{\infty}-
\kappa^{-1}_0)]^{1/2}(2\varepsilon_0V)^{-1/2}$, where
$\kappa_{\infty}$ and $\kappa_0$ are, respectively, high-frequency and
static dielectric constants, $\varepsilon_0$ is permittivity of free
space, $-e$ is the electronic charge, ${q=\mid {\bf q}
\mid}$, and $V$ is volume of the sample.
$\Phi$ is the form-factor, $\Phi(n,m,{\bf q})=\int d^3{\bf
r}\psi^*_n({\bf r})e^{i{\bf qr}}\psi_m({\bf r})$.

This Hamiltonian consists of the electronic subsystem, $H_e$, the bath
$H_{ph}$ and their coupling $H_1$.  A remarkable feature of this
Hamiltonian is the presence of two different components of interaction
operator. This operator can be written as $ H_1=H_1^{(t)}+H_1^{(l)}$.
The transverse part $H_1^{(t)}$ contains that part of $H_1$, in which
the electronic orbital indexes $n,m$ are equal to each other, which
meaning is that electron emits or absorbs phonon without changing
orbital state. The longitudinal term $H_1^{(l)}$ contains terms, in
which the electronic orbital indexes are different, expressing
transition between different orbital states, during emission or
absorption of a phonon. The former term $H_1^{(t)}$ plays a role in
processes of development of a polaron well, after the electron is
placed into an orbital. In the formulation of phonon quantum transport
equation for the distribution function $<b^+_{\bf q}b_{\bf q'}>$ we do
not take into account the process of the development of such a polaron
well in the form of a deformation of the potential in which the
harmonic oscillators of the phonons move in the presence of the charge
carrier. The deformation of the phonon field could be described by a
quantity like the average $<b_{\bf q}>$, for which we are not currently
considering an appropriate transport equation. Therefore, it would not
be unexpected if the term $H_1^{(t)}$ led to a conflicting effect in
the standard phonon transport equation for the density matrix
$<b^+_{\bf q}b_{\bf q'}>$. The transverse term of the electron-phonon
coupling may contribute to this effect already in the first order of
perturbation, while the longitudinal term $H_1^{(l)}$ can contribute to
such an effect only at higher orders.

We need to describe the electron-phonon system with having the phonon
bath passive. This means that energy does not permanently come into or
leave the bath in the long-time limit. The bath energy is related to
the total phonon number given by the statistical distribution
$<b^+_{\bf q}b_{\bf q'}>$. Therefore, it may be meaningful to change
the basis of the electron and phonon states to such a basis, in which
the lattice deformation is already included and the lattice does not
display a tendency to build it up any more, being already accommodated
to the presence of electron. This representation may be achieved upon
performing the well-known Lang-Firsov canonical transformation, which
is known to provide exact diagonalization of the independent boson
model
\cite{mahan}, which in our case corresponds to the Hamiltonian
$H_{IB}=H-H_1^{(l)}$.

We shall show the result of the Lang-Firsov transformation of the
Hamiltonian $H$ and we shall also present the transformation of the
remaining longitudinal part of the coupling $H_1^{(l)}$. We shall
calculate the magnitude of the permanent phonon generation in the
phonon bath within the quantum transport equation based on the full
interaction operator $H_1$. Then we shall show how much the Lang-Firsov
transformation helps to suppress the phonon overheating effect.


With help of canonical Lang-Firsov (LF) transformation $S$ we introduce
new particle operators of electron, $A_n$,  and of phonons, $B_{\bf
q}$, by relations $c_n=SA_nS^+$ and $b_{\bf q}=SB_{\bf q}S^+$. The
operator $S=e^\sigma$, and $\sigma=-\sum_{j,{\bf q}}(A_q\Phi(j,j,{\bf
q})/E_{LO}) (B_{\bf q}+B_{-{\bf q}})A^+_jA_j$. This transformation
diagonalizes exactly the independent boson part of the full
Hamiltonian, $H_{IB}$. Therefore, the full Hamiltonian has now the form
\be
H=\sum_n(E_n-\frac{\alpha_{nn}}{E_{LO}})A^+_nA_n+
\sum_{\bf q}E_{LO}B^+_{\bf q} B_{\bf q}
-\frac{\kappa_{01}}{E_{LO}}A^+_1A_1A^+_0A_0+\tilde{H}_1^{(l)},
\label{hamik}
\ee
where $\kappa_{01}=2\sum_{\bf q} \mid A_q\mid^2 Re [\Phi(0,0,{\bf
q})\Phi(1,1,{\bf q})]$, while the electronic polaron well depth is
$\alpha_{nn}/E_{LO}$, with  $\alpha_{nn}=\sum_{\bf q}
\mid A_q\mid ^2\mid
\Phi(n,n,{\bf q})\mid^2$. $\tilde{H}_1^{(l)}$ is the longitudinal part
of the electron-phonon interaction transformed by the canonical
transformation $S$. This operator comes out from the transformation as
$\tilde{H}_1^{(l)}=V_1+V_2$. Before specifying these two operators, let
us introduce an abbreviation, $\gamma_{j,{\bf q}}=A_q\Phi(j,j,{\bf
q})/E_{LO}$, which is a quantity of the first order in the
electron-phonon coupling $A_q$. The operator $V_2$ reads:
\be
V_2=2\sum_{j,{\bf q}}A_q\Phi(0,1,{\bf q})\gamma_{j,-{\bf q}}A_0^+A_1
e^{\beta} + h.c.,
\ee
where $\beta=\sum_{\bf p}(\gamma_{1,{\bf p}}-\gamma_{0,{\bf p}})(B_{\bf
p }+B^+_{-{\bf p}})$. Let us note that this part of the electron-phonon
interaction is at lest of the second order in the coupling constant
$A_q$. In the lowest order in the electron-phonon coupling constant it
gives a phonon-less electronic level mixing. We shall neglect this
effect. Because in $V_2$ the electron-phonon coupling starts in the
third order of the expansion in powers of the coupling, we shall
neglect $V_2$ in comparison with $V_1$, which reads:
\be
V_1=\sum_{\bf q}A_q\Phi(0,1,{\bf q})(B_{\bf q}-B^+_{-{\bf
q}})A^+_0A_1e^{\beta} + h.c.
\ee
 The effect of the Lang-Firsov transformation has thus
been concentrated in the latter formula into the exponential factor,
which now contains a complicated dependence on the phonon operators.
The advantage of making the LF transformation is also in that we can
further simplify the electron-phonon operator $V_1$. The expansion of
$V_1$ in powers of the coupling starts with the first power of $A_q$.
The lowest order term corresponds to putting $\beta$ equal zero.
Confining ourselves to keeping only the leading order terms in $V_1$,
we estimate the influence of the exponential factor. Namely, we
approximate $e^{\beta}$ by an average of this quantity $ < e^{\beta}>$,
calculated with canonical ensemble at a temperature $T$. Calculating
this average \cite{mahan} we get $< e^{\beta}> =\exp[-\sum_{\bf p}\mid
\gamma_{1,{\bf p}}-\gamma_{0,{\bf p}}\mid^2 (< B^+_{\bf p}B_{\bf
p}> +1/2)]$. Here $< B^+_{\bf p}B_{\bf p}> $ has the meaning of an
average number of phonons with wave vector ${\bf p}$. We shall first of
all assume that this number of phonons is small, comparable with the
factor $1/2$. For the second, we see that the order of the square of
difference of the coupling constants $\gamma_{n,{\bf p}}$ is given
approximately by the  polaron constant \cite{mahan}, which is about
0.07 in GaAs. From these reasons we approximate the factors $e^{\beta}$
by the value of 1. After having made these approximations we can see
that the longitudinal part of the coupling has not changed upon making
the LF transformation.

\begin{figure}[th]
\begin{center}
\includegraphics[width=3in]{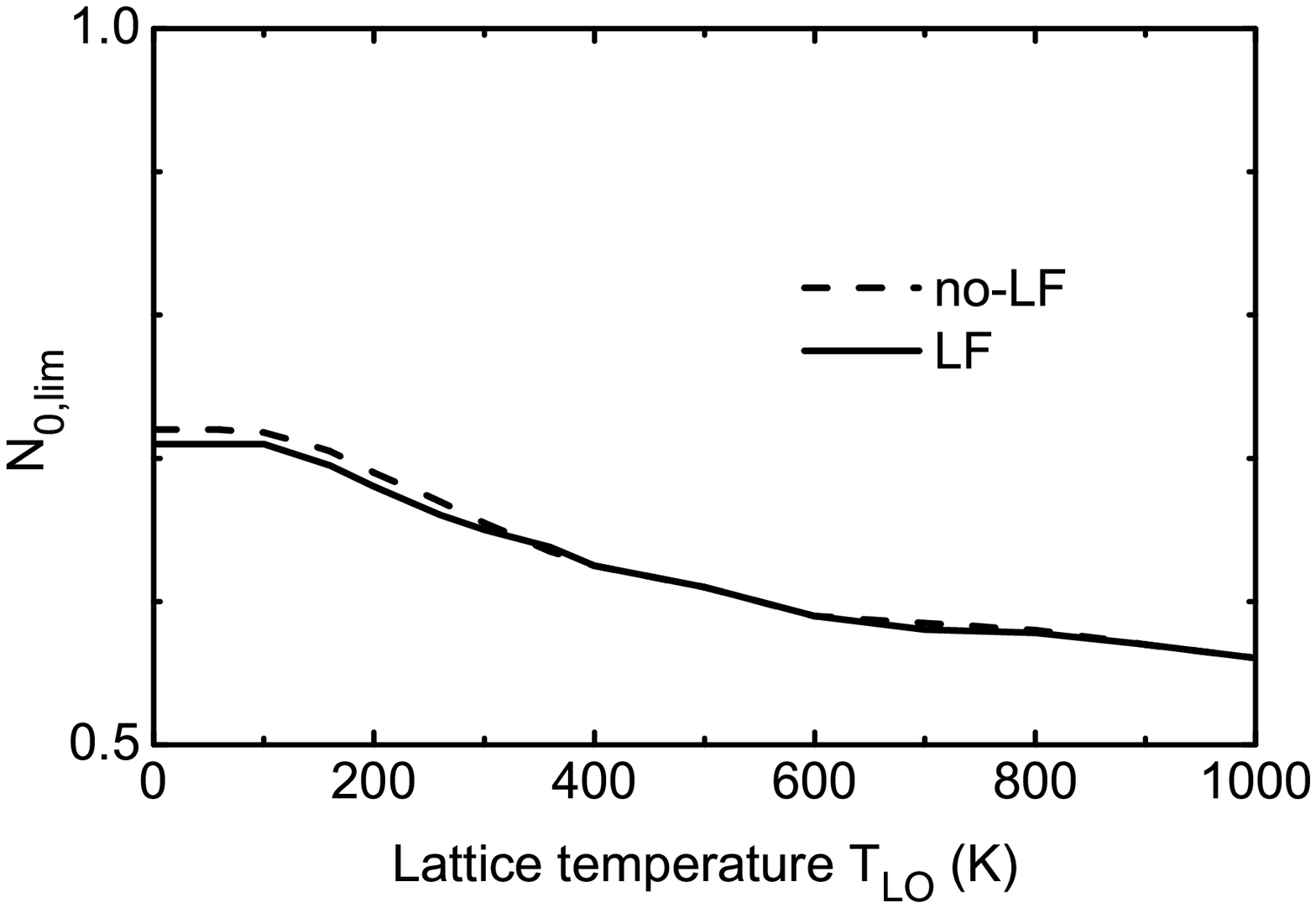}
\caption{Long-time limit (steady state) of occupation $N_0$} of
electronic lower-energy state at given temperature $T_{LO}$ of
LO-phonons. Dashed line is calculated without LF transformation.
\label{N0}
\end{center}

\end{figure}

We also neglect the polaron well depth component of new electron
energy, $\alpha_{nn}/E_{LO}$, in (\ref{hamik}) and the term
proportional to $\kappa_{01}$ in the same equation. These terms are
small on the scale of the electron energy level separations, and the
optical phonon energy, in cases we are interested in. With the above
made approximations, the resulting Hamiltonian of the system, after
making LF transformation, is the operator identical with the original
Hamiltonian $H$ of equation (\ref{original}), with the only change that
the transverse coupling $H_1^{(t)}$ has been dropped. In other words,
when estimating numerically the electronic relaxation in the dressed
electron and phonon basis, the original Hamiltonian (\ref{original})
can be used, with the transverse terms put equal zero. The removal of
the transverse terms from the original Hamiltonian is done simply by
putting the form-factors $\Phi(n,n,{\bf q})$ equal zero for $n=0,1$.

\section{Electron relaxation and phonon heat generation}

In order to demonstrate the overheating effect, we calculate first the
heat generation in the phonon system with using the full original
Hamiltonian (\ref{original}). The kinetic equations for the electron
was derived earlier \cite{japss1998,jaIEEE2004} with help of either
nonequilibrium statistical operator or nonequilibrium Green's functions
\cite{zubarev,LL10}. The derivation of the kinetic equation for the phonon
system can be obtained in the same way. The state of the phonon system
should be generally described with help of the density matrix
$<b^+_{\bf q}b_{\bf q'}>$, or equivalently with help of the Wigner
function. This generality would be good for expressing correctly the
generation of phonons in the area of the quantum dot only. In order to
reduce the computing demands as much as possible, while conserving
reasonably much the predictivity of the model, we reduce somewhat the
complexity of the system of phonon modes. In that sense, in order to
get rid of the off-diagonal elements of the phonon density matrix, we
identify the volume of the basic area of the sample with the volume of
the quantum dot. In addition to this, we formally reduce the number of
the modes, with which the electron interacts, to just a single one,
with phonon annihilation operator $b$, the interaction constant of this
mode with the electron  being chosen such that the coupling of electron
to this single mode provides the same relaxation rapidity as the
original full set of LO phonon modes. We shall publish the
argumentation leading to this simplification elsewhere. The kinetic
equations for the electron  distribution function in the excited state,
$< c^+_1c_1>$, is derived in the self-consistent Born Approximation to
electronic self-energy and in the random-phase approximation to the
phonon self-energy. In the right hand sides of the kinetic equations,
and in the equation for the self-energy, the phonon propagator is
approximated by undressed phonon propagator. Instantaneous collisions
approximation is used. Standard Kadanoff-Baym ansatz is used, which
step is regarded as suitable enough for expressing the long-time limits
of development of the system's statistics. Other details of the
derivation of the kinetic equations will also be described elsewhere.

\begin{figure}[th]
\begin{center}
\includegraphics[width=3in]{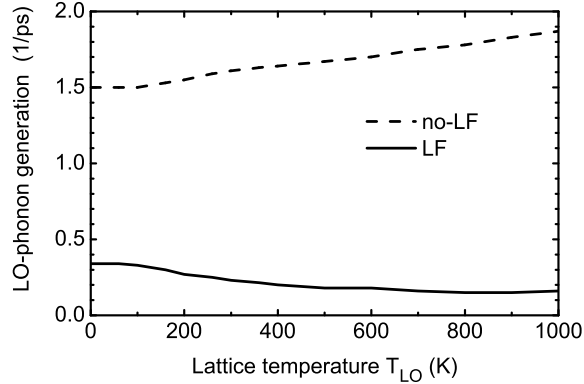}
\caption{Plot of generation rate $d<b^+b>/dt$ of LO-phonons as a function
of temperature $T_{LO}$ of the lattice. The steady-state occupation
$N_0$ of the lower-energy electronic state  is shown in Fig.\
\ref{N0}.}
\label{heating}
\end{center}
\end{figure}

We shall first show the electronic relaxation at a given temperature of
the LO-phonon reservoir. We use the material constants of GaAs, with
energy of LO phonon of 36.2 meV. The lateral size of the cubic dot is
taken to be 16.4 nm. This size corresponds to the electronic energy
level separation of  62.7 meV. The electronic transport equation
calculated with the help of the full Hamiltonian in the original form
(\ref{original}) was solved first to find the electronic distribution
among the two levels at a given lattice temperature. In Fig.\ \ref{N0}
the dashed line shows the lattice temperature dependence of the
occupation $N_0$ of the lower energy electron orbital, at which the
relaxation rate $dN_0/dt$ is zero. In other words, at this $N_0$ the
electronic subsystem reaches the long-time limiting steady state
distribution. Let us remark that we have earlier shown \cite{kralFQMT}
that these values of $N_0$ are well below the values given by the
Fermi-Dirac distribution of electron in the present electronic
two-level system. The dashed curve at Fig.\ \ref{heating} shows the
generation rate of LO phonons $\frac{d<b^+b>}{dt}$ at given phonon
temperature and at the steady state electronic distribution given by
the dashed line in Fig.\
\ref{N0}. One can see that at these conditions the phonon system does
not have a steady state. It permanently heats up with the rather high
speed of about 1.5 to 1.8 phonons per picosecond (the dashed line). We
ascribe this fast heating to the role of the transverse coupling terms
in the original Hamiltonian (\ref{original}), as discussed in an
earlier section. From the above given reasons we go over to a different
representation, using dressed electron and phonon basis, writing
quantum transport equations for electron and the phonon transformed
with help of Lang-Firsov transformation. In this new representation we
use an approximate Hamiltonian, which equals the original operator
(\ref{original}) with the transverse terms dropped. When the same
characteristics of electron and the phonon are calculated, we get the
full lines in both Figs.\
\ref{N0} and
\ref{heating}. We see that in the new basis the phonon heating is
considerably reduced, namely by the factor of about 7, with the present
simple tool based on LF transformation.

Summing up, let us note the remarkable numerical result for the
long-time limit of electronic distribution at given lattice
temperature: this long-time limit practically does not depend on
whether the calculation is done with or without the Lang-Firsov
transformation procedure applied. This helps us to make a conclusion:
the use of the original representation in undressed states, using the
full Hamiltonian (\ref{original}) having the property of the phonon
overheating effect, appears to give us plausible data about the
electronic relaxation in the quantum dot. From this reason, the earlier
made conclusions about the electron relaxation effects
\cite{jaIEEE2004,jaSS2004}, may remain valid. As for the phonon kinetic
equation, the transformation to the dressed basis with LF
transformation helps to reduce the overheating effect considerably.



\noindent
{\small Acknowledgement. This work was supported by institutional
project AVOZ10100520.}

\end {document}